\documentclass[
    ,final            
  ]
  {aipproc}

\layoutstyle{8x11double}

\begin{document}

\title{Evolution of PAH Features from Proto- to Planetary Nebulae}


\author{R. Szczerba}{
  address={N. Copernicus Astronomical Center, Rabia\'nska 8, 87-100 
Toru\'n, Poland}
}

\author{N. Si\'odmiak}{
  address={N. Copernicus Astronomical Center, Rabia\'nska 8, 87-100 
Toru\'n, Poland}
}

\author{C. Szyszka}{
  address={Toru\'n Centre for Astronomy of the NCU, Gagarina 11, 87-100 
Toru\'n, Poland}
}

\begin{abstract}
With the aim to investigate the overall evolution of UIR band features with
hardening of UV radiation (increase of the star's effective temperature) we
have analysed ISO spectra for 32 C-rich stars: 20 proto-planetary nebulae and 
12 planetary nebulae with Wolf-Rayet central stars. In this contribution we 
discuss variations in the peak position of UIR bands among analysed objects,  
and demonstrate that variations in the ``7.7'' to 11.3\,$\mu$m flux ratio are 
correlated with the effective temperature (probably due to an increase 
of the ionization state of their carriers). 
\end{abstract}

\maketitle


\section{Introduction}

The presence of unidentified infra-red (UIR) bands around 3.3, 6.2, ``7.7'', 
8.6 and 11.3 $\mu$m in many astronomical objects is commonly attributed to 
polycyclic aromatic hydrocarbon (PAH)  molecules which are 
excited by UV photons (see e.g. Li\,2004). After absorption of a
single UV photon, energy is redistributed over the molecule and the molecule 
``cools down'' through photon emission in the UIR bands.
Possibly, the observed features are due to a complex mixture of ionized 
and neutral PAH or PAH-related molecules of different sizes. 
Therefore, the overall appearance of the features (shape, peak position and 
strength) are determined by the present physical conditions, but also by the
formation and evolution of their carriers.  

PAHs are thought to be formed in the outflows from C-rich asymptotic giant 
stars (AGB). However, there is no direct evidence of the presence of these 
large molecules in circumstellar envelopes of C-rich AGB 
stars\footnote{Recently, Boersma et al.\,(2005) reported the detection of the 
UIR bands in the Infrared Space Observatory spectrum of the AGB carbon star TU 
Tau: a binary star which has a blue companion.}. The most possible explanation 
of UIRs undetection is the lack of UV photons in surroundings of AGB stars. 

Recently, Peeters et al.\,(2002) and van Diedenhoven et al.\,(2004) have 
investigated variations of the UIR features in a large variety of sources 
including star forming regions, H\,II regions, Herbig Ae Be stars and galaxies 
as well as evolutionary advanced stellar objects. However, their sample included
only 7 post-AGB objects and 4 planetary nebulae (PNe) with 
[WR]\footnote{Wolf-Rayet planetary nebulae have central stars which are hydrogen
poor and lose mass at higher rates than normal central stars, but their 
chemical composition is similar to that for normal PNe. ISO data have 
shown that in this group of objects both forms of dust (C-rich: responsible for
UIR bands and O-rich: responsible for crystalline silicate features) are 
present (Waters et al.\,1998, Cohen et al.1999).} central stars. Therefore, to 
investigate variations (and possible evolution) of UIR features in sources with
fairly well established physical conditions we have determined parameters of 
UIR band profiles for the largest available sample of post-AGB objects and [WR] 
PNe using Infrared Space Observatory (ISO, see Kessler et al.\,1996) data.

\section{Sample and the UIR features}
We have searched the ISO Data Archive for Short Wavelength Spectrometer (SWS, 
see de Graauw et al.\,1996) data taken with AOT\,01 for about 330 post-AGB 
objects compiled recently by Si\'odmiak\,(2005). SWS\,01 data are available for
65 sources and UIR bands are present in spectra for 20 of them. The sample of 
[WR] PNe 
includes sources which were discussed by Szczerba et al.\,(2001), however, we 
excluded 4 objects which do not show UIR bands or have SWS\,01 data which do 
not allow to make a quantitative analysis of their UIR bands. 

The ISO Spectral Analysis Package (ISAP\,2.1)\footnote{The ISO Spectral 
Analysis Package (ISAP) is a joint development by the LWS and SWS Instrument 
Teams and Data Centers. Contributing institutes are CESR, IAS, IPAC, MPE, RAL
and SRON.} was used to process and analyze SWS\,01 spectra (OLP version 10.1) 
for all sources from our sample. During data reduction bad data were removed 
and spectra were rebinned to a constant resolution of 300. Small memory effects 
were 
smeared out by direct averaging across up and down scans, but in case of larger
memory effects, the up and down scans were averaged separately to investigate
changes in the UIR features. The parameters of the UIR band profiles were 
determined by defining local continua (polynomial of order 1) and fitting a 
single Gaussian to the features at 3.3, 6.2, ``7.7'', 8.6 and 11.3\,$\mu$m, 
separately. The ``7.7''\,$\mu$m complex is treated similarly in spite of
the evidence that (in many cases) this band is composed of at least two variable
components (see Peeters et al.\,2002 and references therein). In addition, we 
have used a single baseline underlying the ``7.7'' and 8.6\,$\mu$m features, 
which passes through points from $\sim $6-7\,$\mu$m up to $\sim$9-10\,$\mu$m. 
Other ways of decomposing the UIR bands will yield slightly different results. 
However, these differences will affect all sources in a systematic way and the 
variations discussed here will remain.

We present in Table\,1 our sample (IRAS name in column (2), galactic coordinates
- column
(3), the effective temperature, T$_{\rm eff}$ - column (4) - if available) and 
the obtained UIR band parameters (peak position, $\lambda_{\rm c}$\,[$\mu$m], 
and the integrated fluxes of the Gaussian fits, F\,[W/cm$^2$]) for the 3.3, 
6.2, ``7.7'', 8.6 and 11.3\,$\mu$m features (columns (5), (6), (7), (8) and 
(9), respectively). The shapes of the band profiles will be
discussed in the full version of the paper (Szczerba et al. in preparation).

\begin{table}[th!]
\begin{tabular}{lllrlllll}
\hline
 \tablehead{1}{l}{b}{No \\  \\ (1)} &
 \tablehead{1}{l}{b}{Object name \\ \\ (2)} &
 \tablehead{1}{l}{b}{l~~~~~~~~~~~~~~b \\  \\ (3)} &
 \tablehead{1}{c}{b}{T$_{\rm eff}$ [K] \\  \\ (4)} &
 \tablehead{1}{c}{b}{3.3\,$\mu$m \\ ${\lambda}_{\rm c}\,[\mu$m]  ; F\,[W/cm$^2$] \\ (5)} &
 \tablehead{1}{c}{b}{6.2\,$\mu$m \\ ${\lambda}_{\rm c}$ ; F \\ (6)} &
 \tablehead{1}{c}{b}{``7.7''\,$\mu$m \\ ${\lambda}_{\rm c}$ ; F \\ (7)} &
 \tablehead{1}{c}{b}{8.6\,$\mu$m \\ ${\lambda}_{\rm c}$ ; F  \\ (8)} &
 \tablehead{1}{c}{b}{11.3\,$\mu$m \\\ ${\lambda}_{\rm c}$ ; F  \\ (9)} \\
\hline
\noalign{\smallskip}
\multicolumn{7}{l}{\bf proto-planetary nebulae:} \\
\noalign{\smallskip}
1  & IRAS 19480$+$2504 & 061.84 $-$00.56  &           &                   &                  &                  &8.502 ; 7.684e-18  &11.628 ; 3.626e-18  \\
2  & IRAS 20000$+$3239 & 069.68 $+$01.16  &  5500.    &3.294 ; 3.098e-19  &6.277 ; 1.065e-18 &8.110 ; 1.272e-17 &                   &11.513 ; 4.321e-18  \\
3  & Egg Nebula        & 080.17 $-$06.50  &  6500.    &3.293 ; 4.056e-19  &6.291 ; 9.155e-18 &8.172 ; 7.294e-17 &                   &11.357 ; 1.844e-17  \\
4  & IRAS 22272$+$5435 & 103.35 $-$02.52  &  5750.    &3.275 ; 3.850e-19  &6.252 ; 6.431e-18 &7.967 ; 3.766e-17 &8.631 ; 4.377e-18  &11.346 ; 1.615e-17  \\
5  & IRAS 22574$+$6609 & 112.04 $+$05.96  &  5500.    &                   &6.254 ; 1.194e-18 &7.907 ; 1.076e-17 &8.517 ; 4.378e-18  &11.379 ; 5.350e-18  \\
6  & IRAS 23304$+$6147 & 113.86 $+$00.59  &  6750.    &                   &                  &7.963 ; 1.222e-17 &                   &11.371 ; 3.989e-18  \\
7  & IRAS 01005$+$7910 & 123.57 $+$16.59  & 21000.    &3.277 ; 3.808e-19  &6.228 ; 1.649e-18 &7.565 ; 9.873e-18 &8.625 ; 1.293e-18  &11.258 ; 1.275e-18  \\
8  & IRAS Z02229$+$6208& 133.73 $+$01.50  &  5500.    &                   &6.279 ; 5.352e-18 &7.920 ; 1.625e-17 &8.650 ; 6.917e-18  &11.378 ; 1.365e-17  \\
9  & IRAS 04395$+$3601 & 166.45 $-$06.53  & 25000.    &3.259 ; 5.649e-19  &                  &                  &                   &                    \\
10 & IRAS 05341$+$0852 & 196.19 $-$12.14  &  6500.    &                   &6.284 ; 8.420e-19 &7.833 ; 1.180e-17 &                   &11.333 ; 1.501e-18  \\
11 & IRAS 07134$+$1005 & 206.75 $+$09.99  &  7250.    &                   &                  &7.766 ; 7.799e-18 &                   &11.316 ; 5.813e-18  \\
12 & IRAS 06176$-$1036 & 218.97 $-$11.76  &  7500.    &3.294 ; 6.110e-17  &6.267 ; 2.067e-16 &7.814 ; 4.893e-16 &8.665 ; 9.627e-17  &11.264 ; 9.085e-17  \\
13 & IRAS 10158$-$2844 & 266.85 $+$22.93  &  7600.    &3.293 ; 4.493e-18  &6.269 ; 2.110e-17 &7.821 ; 4.416e-17 &8.659 ; 9.063e-18  &11.272 ; 8.464e-18  \\
14 & IRAS 13416$-$6243 & 308.99 $-$00.73  &           &3.295 ; 1.002e-18  &6.292 ; 4.080e-18 &8.192 ; 3.504e-17 &                   &11.417 ; 4.269e-18  \\
15 & IRAS 13428$-$6232 & 309.16 $-$00.59  &           &                   &6.255 ; 1.889e-18 &7.756 ; 5.042e-18 &8.625 ; 1.017e-18  &11.305 ; 1.308e-18  \\
16 & IRAS 14316$-$3920 & 323.77 $+$19.10  &  6750.    &3.293 ; 1.100e-18  &6.301 ; 7.204e-18 &                  &                   &11.339 ; 1.045e-18  \\
17 & IRAS 16279$-$4757 & 336.14 $+$00.09  &  4900.    &3.288 ; 1.641e-18  &6.235 ; 1.221e-17 &7.720 ; 2.798e-17 &8.595 ; 2.419e-18  &11.265 ; 8.426e-18  \\
18 & IRAS 16594$-$4656 & 340.39 $-$03.29  & 12000.    &3.289 ; 9.949e-19  &6.244 ; 9.624e-18 &7.782 ; 3.053e-17 &8.580 ; 9.373e-18  &11.271 ; 4.024e-18  \\
19 & IRAS 17311$-$4924 & 341.41 $-$09.04  & 21000.    &                   &6.266 ; 1.094e-18 &7.822 ; 3.436e-18 &8.629 ; 6.672e-19  &11.280 ; 7.724e-19  \\
20 & IRAS 17347$-$3139 & 356.80 $-$00.06  & 15000.    &3.295 ; 4.315e-19  &6.267 ; 5.226e-18 &7.961 ; 2.032e-17 &8.653 ; 4.314e-18  &11.409 ; 2.433e-18  \\
\noalign{\smallskip}
\multicolumn{7}{l}{\bf [WR] planetary nebulae:} \\
\noalign{\smallskip}
1  & IRAS 18129$-$3053 & 001.59 $-$06.72  & 35000.    &3.290 ; 7.276e-19  &6.275 ; 8.078e-18 &7.854 ; 9.948e-18 &8.658 ; 8.758e-19  &11.215 ; 1.213e-18  \\ 
2  & IRAS 17262$-$2343 & 002.43 $+$05.85  &151000.    &3.305 ; 9.031e-19  &                  &                  &                   &11.317 ; 4.227e-18  \\
3  & IRAS 18240$-$0244 & 027.68 $+$04.26  & 65000.    &3.292 ; 1.663e-18  &6.257 ; 8.636e-18 &7.940 ; 3.510e-17 &8.631 ;1.072e-17   &11.298 ; 4.524e-18  \\
4  & IRAS 19327$+$3024 & 064.79 $+$05.02  & 47000.    &3.290 ; 5.960e-18  &                  &7.826 ; 1.054e-16 &8.576 ; 3.865e-17  &11.277 ; 2.967e-17  \\ 
5  & IRAS 20119$+$2924 & 068.35 $-$02.74  & 77000.    &3.315 ; 5.472e-19  &                  &                  &8.731 ; 2.482e-18  &11.240 ; 2.540e-18  \\ 
6  & IRAS 00102$+$7214 & 120.02 $+$09.87  & 78000.    &3.295 ; 3.046e-19  &                  &7.792 ; 3.043e-18 &8.643 ; 9.087e-19  &11.239 ; 9.369e-19  \\ 
7  & IRAS 04215$+$6000 & 146.79 $+$07.60  & 31000.    &                   &6.229 ; 3.572e-18 &7.733 ; 8.990e-18 &8.758 ; 2.035e-18  &11.285 ; 1.197e-18  \\ 
8  & IRAS 07027$-$7934 & 291.38 $-$26.29  &           &3.290 ; 1.554e-18  &6.286 ; 1.419e-17 &7.967 ; 4.634e-17 &8.631 ; 1.074e-17  &11.300 ; 9.212e-18  \\ 
9  & IRAS 13501$-$6616 & 309.11 $-$04.40  &           &3.234 ; 8.292e-19  &                  &                  &                   &11.239 ; 3.841e-18  \\
10 & IRAS 14562$-$5406 & 321.05 $+$03.99  & 30000.    &3.291 ; 6.416e-18  &6.240 ; 4.086e-17 &7.914 ; 1.351e-16 &8.589 ; 4.295e-17  &11.293 ; 1.928e-17  \\ 
11 & IRAS 15559$-$5546 & 327.19 $-$02.20  &           &3.288 ; 1.007e-18  &                  &7.781 ; 1.887e-17 &8.633 ; 5.031e-18  &11.254 ; 4.415e-18  \\ 
12 & IRAS 17047$-$5650 & 332.92 $-$09.91  & 32000.    &3.290 ; 7.368e-18  &6.258 ; 4.816e-17 &7.849 ; 1.919e-16 &8.587 ; 4.499e-17  &11.298 ; 3.127e-17  \\
\hline
\end{tabular}
\caption{Central wavelengths and fluxes of UIR bands for sample of C--rich proto--planetary and [WR] planetary nebulae.}
\label{tab:flux}
\end{table}

\section{Discussion}
\begin{figure}
\includegraphics[width=\columnwidth]{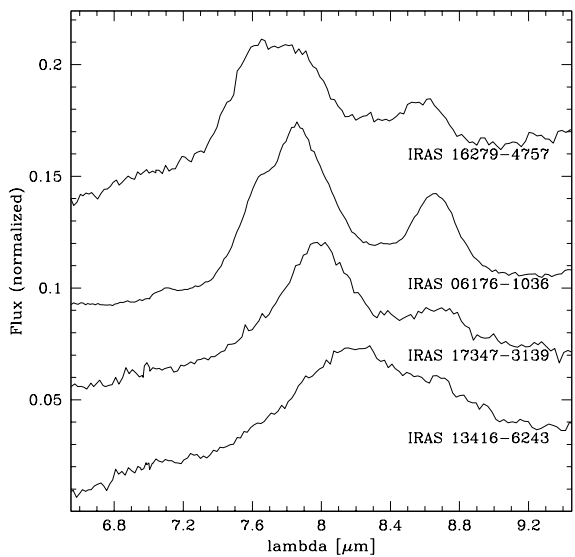} 
\caption{The normalized spectra of four post-AGB objects from our sample 
showing variations in the ``7.7'' $\mu$m peak shape and position.} 
\label{Fig1}
\end{figure}

The ``7.7'' $\mu$m band is composed of at least two sub-peaks and in some cases
the main peak shifts up to 8\,$\mu$m or more (see Peeters et al.\,2002 and 
references therein). In Fig.\,1 we show spectra of four post-AGB sources 
selected from our sample, which are ordered according to the peak wavelength of 
the ``7.7'' $\mu$m complex. The whole range of possible peak shapes and 
positions are seen among post-AGB sources.
  
Peeters et al. (2002) and van Diedenhoven et al.\,(2004) classified objects, independently for each UIR band, into classes 
based on the band profile and peak position. They noted that the derived classes are directly linked with each other. 
Fig.\,2 shows such a correlation between the ``7.7'' and 6.2\,$\mu$m peak 
position for our sample of proto-PNe and PNe. In general,
sources with a 6.2 $\mu$m feature peaking at longer wavelengths show also a ``7.7'' $\mu$m complex shifted toward the red.
There is no sharp ``jump'' between their classes but rather a continuous transition in the peak position for both features It is 
possible that there is an abrupt ``jump'' to the band peaking at 8.1-8.2\,$\mu$m. However, sources with the peak wavelength of 
the ``7.7'' complex shifted above 8 $\mu$m have a 6.2 $\mu$m band peaking at the longest wavelengths in accordance with the
overall correlation seen in Fig.\,2. Note, that there is no separation between post-AGB objects (circles) and [WR] PNe 
(triangles) on this diagram. 

\begin{figure}
\includegraphics[width=\columnwidth]{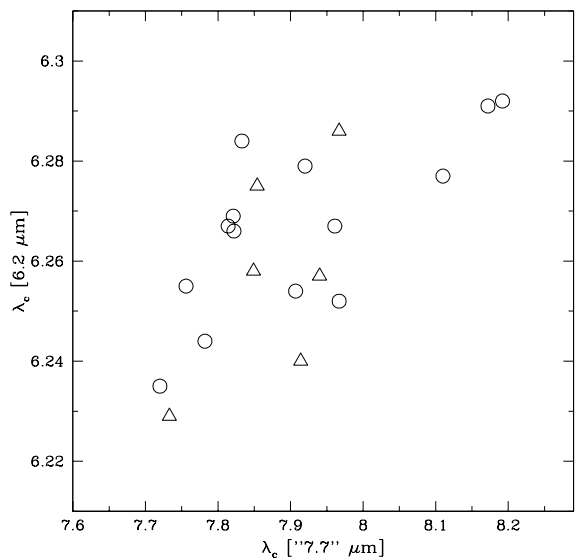} 
\caption{Peak position of the ``7.7'' $\mu$m complex versus that of the 6.2 $\mu$m band. Open circles denote post-AGB objects and open triangles mark [WR] PNe.}
\label{Fig2}
\end{figure}

The peak of the 8.6\,$\mu$m band remains well confined between 8.57 and 8.67 $\mu$m (with two exceptions) and does not show any
correlation with the peak position of the ``7.7'' $\mu$m complex. Note, however, that there is no 8.6 $\mu$m band for objects which
have ``7.7'' band peak shifted above $\sim$8 $\mu$m. The two exception are: IRAS\,22574$+$6609 (proto-PN no. 5 in 
Table 1.) with $\lambda_{\rm c}$ [8.6\,$\mu$m] = 8.52 $\mu$m, and IRAS\,04215$+$6000 ([WR] PN no. 7) with
$\lambda_{\rm c}$ [8.6\,$\mu$m] = 8.76 $\mu$m. In the latter case the peak position is uncertain due to low S/N of the ISO
spectrum. 

The peak of the 3.3 $\mu$m band is even more confined. It is located between 3.287 
and 3.296\,$\mu$m with the exception of two post-AGB objects: IRAS\,22272$+$5435 
(no.\,4) which have $\lambda_{\rm c}$ [3.3\,$\mu$m] = 3.275\,$\mu$m, 
and IRAS\,01005$+$7910 (no.\,7) with $\lambda_{\rm c}$ [3.3\,$\mu$m] = 
3.277\,$\mu$m. On the other hand, the peak position of the 11.3\,$\mu$m band is not so 
``stable'' ($\Delta${$\lambda_{\rm c}$ [11.3\,$\mu$m]} $\approx$0.2\,$\mu$m).
Note that $\lambda_{\rm c}$ [11.3\,$\mu$m] correlates with the ``7.7''\,$\mu$m 
band peak in a similar way as shown in Fig.2. 

\begin{figure}
\includegraphics[width=\columnwidth]{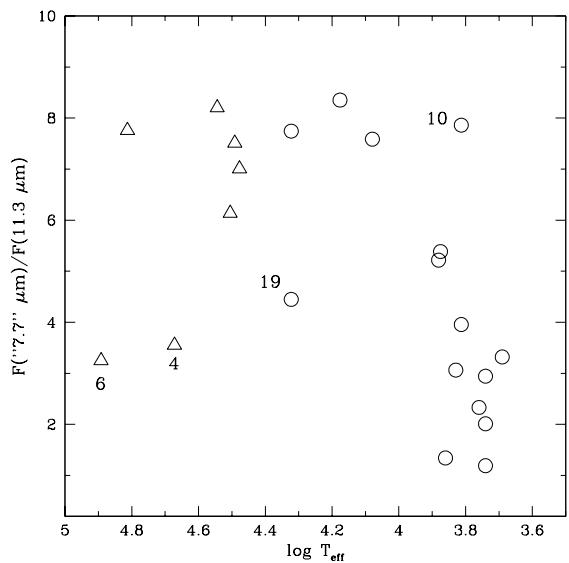} 
\caption{Flux ratio of the ``7.7'' $\mu$m complex and the 11.3 $\mu$m band. 
Meaning of symbols is the same as in Fig.\,2. Numbers from Table\,1 identify objects 
for which position on the diagram deviates from the general trend.}
\label{Fig3}
\end{figure}

The ``7.7'' $\mu$m complex (if present) is one of the strongest among the UIR 
bands. Its strength depends critically on the ionization state of the emitting 
PAH molecules (see discussion in Peeters et al.\,2002). During evolution from 
AGB to PNe the 
stellar effective temperature increases steadily and we can expect 
that PAH molecules formed on the AGB will become ionized and the
``7.7'' $\mu$m to 11.3 $\mu$m flux ratio (which decreases after ionization 
of the PAH molecules) will become higher. Fig.\,3 shows this flux ratio
as a function of the stellar effective temperature. If one discards [WR]\,PNe no.
4 and 6 and proto-PN no.19, the general tendency is an increase of this flux 
ratio with increase of T$_{\rm eff}$, levelling off at about 
log(T$_{\rm eff}$)$\sim$4.2. The exceptions are: post-AGB object no.19 - IRAS\,17311$-$4924 and [WR] PNe no.4 - IRAS\,19327$+$3024 and no.6 - IRAS\,00102$+$7214.
These exceptions can not be rather explained by the uncertainties in their spectra.
It seems also that IRAS\,05341$+$0852 (no.10) has slightly too high 
F(``7.7'' $\mu$m)/F(11.3 $\mu$m) for so low stellar temperature. 

\section{Conclusions}
By investigation of UIR band features in 20 post-AGB objects and 12 [WR] PNe (a 
sample more limited from a physical point of view than the sample analysed by 
Peeters et al.\,2002 and van Diedenhoven et al.\,2004 which includes star 
forming regions, H\,II regions, Herbig Ae Be stars, galaxies as well as 
proto-PNe and PNe) we have shown that there
are clear variations in the UIR bands parameters resulting (possibly) from PAH 
modification (chemical or/and physical) due to the circumstellar shell and/or 
star evolution. We have shown the existence of a correlation between UIR band 
flux ratios and the star's temperature. Possibly the ionization of the UIR band 
carriers is responsible for variation of F(``7.7'' $\mu$m)/F(11.3 $\mu$m) with 
the hardening of the radiation field. Such a conclusion could be reached because
the effective temperature can be determined for proto-PNe and PNe. In the
heterogenous sample of Peeters et al.\,(2002) there is no easy way to determine
the physical parameters in a consistent manner.

\begin{theacknowledgments}
This work has been partly supported by grant 2.P03D.017.25 of the Polish 
State Committee for Scientific Research.
\end{theacknowledgments}

\bibliographystyle{aipproc}   

\end{document}